\documentclass[letter]{jpconf}
\usepackage{graphicx}
\begin{document}
\title{ATLAS Forward Detectors and Physics}

\author{Nitesh Soni
\\ on behalf of the ATLAS collaboration}
\address{Centre for Particle Physics, University of Alberta, Edmonton, Alberta, T6G~2G7, Canada}

\ead{nsoni@phys.ualberta.ca}

\begin{abstract}
In this communication I describe the ATLAS forward physics program and the detectors, LUCID, ZDC and ALFA that have been designed to meet this experimental challenge. In addition to their primary role in the determination of ATLAS luminosity these detectors - in conjunction with the main ATLAS detector - will be used to study soft QCD and diffractive physics in the initial low luminosity phase of ATLAS running. Finally, I will briefly describe the ATLAS Forward Proton (AFP) project that currently represents the future of the ATLAS forward physics program.
\end{abstract}
\section{Introduction}
The ATLAS central detector~\cite{atlas-detector} consists of an inner tracking detector ($|\eta| < 2.5$), electromagnetic and hadronic calorimeters ($|\eta| < 4.9$) and the muon spectrometer ($|\eta| < 2.7$). In addition, ATLAS is also equipped with the LUCID~\cite{lucid}, ZDC~\cite{zdc} and ALFA~\cite{alfa} detectors which partially cover the forward rapidity region.

The cross-sections for elastic and diffractive production are large. At the centre-of-mass energy of 14 TeV the elastic cross-section is estimated to be 25 - 30 mb. The cross-section for single and double diffraction are estimated 10 - 15 mb. Thus, elastic and diffractive processes account for roughly half of the $p-p$ total cross-section of $\sim 100$ mb. Thus only modest luminosity is required to study these processes. This is fortuitous since event pile-up resulting from higher luminosity running will tend to destroy the rapidity gap signature of these forward physics processes.

\section{The Forward Detector System in ATLAS}
\subsection{LUCID: Luminosity measurement using Cerenkov Integrating Detector}
 LUCID is composed of two modules located at $\pm 17$ m from the interaction point that provide a coverage $5.5 < |\eta| < 5.9$ for charged particles. Each LUCID detector is a symmetric array of 1.5 m long polished aluminium tubes that surrounds the beam-pipe and points toward the ATLAS interaction point (IP). This results in a maximum of Cerenkov emission from charged particles from the IP that traverse the full length of the tube. Each tube is 15 mm in diameter and filled with C4F10 gas maintained at a pressure of 1.2 - 1.4 bar giving a Cerenkov threshold of 2.8 GeV for pions and 10 MeV for electrons.  The Cerenkov light emitted by the particle traversing the tube has a half-angle of $3^{o}$ and is reflected an average 3 - 4 times before the light is measured by photomultiplier tubes which match the size of Cerenkov tubes. The fast timing response (a few ns) provides the unambiguous measurements of individual bunch-crossings. LUCID is sitting in the high radiation area that is estimated to receive a radiation dose of $\sim 7$ Mrad per year at maximum luminosity ($10^{34} \rm cm^{-2}s^{-1}$).   

LUCID is a relative luminosity detector and during the initial period of LHC operation, the absolute calibration would come from the LHC machine parameters allowing the luminosity to be determined to a precision of  $\sim 10 - 20$\%. After an initial period of LHC running $W/Z$ boson counting can be used, as the production cross sections are known well enough to allow and absolute luminosity calibration to 5 - 8\% accuracy. QED processes such as exclusive muon pair production via two photon exchange can be calculated to be better than 1\% providing another physics based calibration. However, the rates of such processes are quite low and their experimental acceptance and detection efficiency are difficult to estimate accurately.  

The final absolute luminosity calibration will be determined to a precision of a few percent using elastic proton-proton scattering in the Coulomb Nuclear Interference (CNI) region covered by the ALFA detector. This method requires special low luminosity high beta runs and consequently it is unlikely that this source of calibration will be available in initial LHC running. 

\subsection{ZDC: Zero Degree Calorimeter}

The Zero Degree Calorimeters (ZDCs) provide coverage of the region $|\eta| > 8.3$ for neutral particles. They reside in a slot in the TAN (Target Absorber Neutral) absorber, which would otherwise contain copper shielding. The ZDC is located at $\pm 140$m from the interaction point, at a place where the straight section of the beam-pipe divides into two independent beam-pipes. There will be four ZDC modules installed per arm: one electromagnetic (EM) module and three hadronic modules. Each EM module consists of 11 tungsten plates, with their faces perpendicular to the beam direction. The height of these plates is extended in the vertical direction with 290mm long steel plates. Two types of quartz radiator are used: vertical quartz strips for energy measurement and horizontal quartz rods which provide position information. At present only hadronic modules are installed. The EM module will be installed once the LHCf project has completed data taking.

\subsection{ALFA: Absolute Luminosity For ATLAS}

The Roman-pot spectrometers are located $\pm 240$m away from the interaction point (IP). There will be two Roman pot stations separated by four meters on either side of the IP. The main requirements on the ALFA scintillating fibre detectors that will be housed in the Roman pots are: a spatial resolution of about $30 \mu $m; no significant inactive region; minimal sensitivity to the radio frequency noise from the LHC beams; and, ability to operate in the vacuum maintained in the Roman pots. At the beginning of the run, the ALFA detectors are in withdrawn position far from the beam. After the beam has stabilized, the detectors are moved back to within 1.5mm of the beam. Elastic and diffractive protons deflected from the beam pass through arrays of scintillating fibre trackers (20 $\times$ 64 fibres in each array), which measure the distance of the proton to the beam. 

Traditionally, the absolute luminosity at hadron colliders has been determined via elastic scattering at small angles.  ATLAS also pursues this approach with the ALFA detector. The extremely small angles ($3 \mu \rm rad$) needed to make these measurements are smaller than the nominal beam divergence. So special beam conditions e.g. high-beta ($\beta ^{*}$) optics in combination with reduced beam emittance, are required. ALFA will be used to determine the absolute luminosity via elastic scattering at small angles in the Coulomb-nuclear interference region. 

\section{Forward Physics Using Initial Data}

\subsection{Single Diffractive di-jet Production}
Single diffractive (SD) can be tagged by identifying the rapidity gap, by requiring that the forward detector system register little hadronic activity. The ATLAS forward calorimeter (FCAL), LUCID and the ZDC can be utilized as part of a rapidity gap requirement for the SD analysis. Di-jet production by SD should be measurable with $\sim 100\rm pb^{-1}$ of the data, corresponding to around 1.5 years of data acquisition at ${\cal L} = 10^{31}\rm cm^{-2} s^{-1}$. The cross-section for SD di-jet production is predicted by the POMWIG event generator~\cite{fp-ref1} to be 3.6 (0.20)$\mu$b for $\zeta < 0.1$ for jet transverse energy ($E_{T}$) greater than 20 (40)GeV, where $\zeta$ is the fractional momentum lost by proton during the interaction. Di-jet production permits a study of factorization breaking in diffractive events. Additional soft interactions and multiple parton-parton scattering during the $pp$-interaction reduce the observed cross-section for diffractive processes at hadron colliders, with respect to the predicted cross-section obtained from diffractive parton distribution function measured at HERA.

\subsection{Central Exclusive Di-jet Production}
The Central Exclusive Production (CEP) is defined as the process $pp \to p\phi \bar p$, where all of the energy lost by proton goes into the production of a hard central system, $\phi$. Thus the final state consists of two outgoing protons, a hard central system and no other activity. The CEP allows direct access to quantum numbers of $\phi$ and has the direct relation of the scattered protons energy loss to the central mass $\rm M_{X}$ with clean azimuthal correlation of the both scattered protons. At certain scenario it could provide a clean Higgs discovery channel \cite{higgs-cep}. 

The di-jet cross section is predicted by the ExHuME event generator~\cite{exhume} to be approximately 8nb for a minimum jet transverse energy of 20GeV. It should be noted that the current measurements of CEP by the CDF collaboration ~\cite{cdfCEP} are in good agreement with the theoretical predictions that form the physics basis of the ExHuME generator. Given the large pre-scale on low $E_{T}$ jets, one would expect approximately 100 events in 100pb$^{-1}$ of data. It is necessary to reduce the Level 1 (L1) pre-scaling to obtain the good measurement of the CEP. It may be possible to do this in ATLAS by exploiting the clean nature of the exclusive event by requiring a rapidity gap in the L1 trigger (using LUCID, ZDC), in conjunction with a triggered jet.

\subsection{Soft Single Diffraction Studies}
The single diffractive (SD) is characterized by a centrally produced system separated by a rapidity gap, or lack of hadronic activity, from an outgoing proton. In SD exchange the outgoing proton can be tagged and measured during special LHC runs by  the ALFA detectors. However, the low luminosity means that only soft SD processes can be studied, in particular the forward proton spectrum at low $\zeta$. The acceptance is  $\sim $ 50\% (10\%) for $\zeta \sim$ 0.01 (0.1). It is expected that at a luminosity of $10^{27} \rm cm^{-2}s^{-1}$ there would be 1.2 to 1.8 million events recorded in 100 hours of data acquisition. The $\zeta$ measurement resolution is approximately 8\% for $\zeta = 0.01$, falling to 2\% for $\zeta = 0.1$.  

\section{The AFP Project - Forward Physics at High Luminosity}

The aim of the  ATLAS Forward Proton (AFP)~\cite{afp} project is to install spectrometers at $\pm 220$m  and $\pm 420$m  from the interaction point of ATLAS. In exclusive central production processes  
where the  incident protons remain intact, the precise measurement of fractional momentum loses ($\zeta _{1}, \zeta _{2}$) can be used to determine the mass of the central system  with great accuracy, using the relation $M^2 = s \zeta _{1} \zeta _{2}$ where $s$ is  square of the center of mass energy. The $\zeta$ acceptance of the AFP detectors at 220m and 420m is $0.01 < \zeta < 0.2$ and $0.002 < \zeta < 0.02$, respectively. This implies a mass acceptance of the central system spanning the range from 80 GeV to masses in excess of 1 TeV.

 AFP project opens up a possibility of searching for new physics in CEP processes, such as Higgs boson production in SM, MSSM and NMSSM~\cite{mssm}. Because of the limited available space at 420m, the traditional Roman Pot technique cannot be used. Instead, the AFP group have opted  to employ so called a Hamburg Movable Beam pipe  system to deploy the AFP forward spectrometers.  In order to achieve a good acceptance and mass resolution, 3D silicon edgeless technology has been chosen for the spectrometers, where the 3D silicon sensors have rectangular pixels of dimensions 50 microns by 400 microns. It is envisaged that the pile-up background can be handled by using ultra precise time-of-flight (ToF) detectors to differentiate the vertex of interest from the vertices of the pile-up events by measuring, the arrival time of the two deflected protons in the ToF detectors with a precision of $\sim$10ps.  There are two approaches to ToF measurement currently being studied. Both of these approaches utilize Cerenkov detectors readout by Microchannel Plate (MCP) photomultipliers. One of these approaches utilizes a gas Cerenkov radiator (GASTOF), whilst the other  employs a fused-silica radiator (QUARTIC). 

\section{Conclusions}
The luminosity monitor LUCID, calibrated by the ALFA detector, will allow the luminosity delivered to ATLAS to be determined to better than 5\% accuracy. The ZDC will measure forward spectators for heavy ion collisions and provide trigger and centrality measurements. It will also provide a luminosity measurement and, measure forward particle production for MC tuning. Low luminosity forward physics topics include: elastic scattering using ALFA; SD forward proton spectrum (ALFA); single diffractive di-jet and $W$ production and di-jets from Double Pomeron Exchange (DPE) and CEP (with rapidity gap vetoes from FCAL, LUCID, ZDC). At high luminosities the ATLAS Forward Proton (AFP) project aims to deploy proton taggers at  $\pm 220$m and $\pm 420$m in order to obtain access to a rich new vein of CEP physics, that includes SM/MSSM/NMSSM Higgs boson studies, $W$ pair production, slepton production and gluino pair production, etc.


\section*{References}
\begin{thebibliography}{9}
\bibitem{atlas-detector} G. Aad {\it et al.}, (The ATLAS Collaboration), JINST 3 S08003 (2008). 
\bibitem{lucid} G. Aad {\it et al.}, (The ATLAS Collaboration), JINST 3 S08003, {\bf 206} (2008). 
\bibitem{zdc} G. Aad {\it et al.}, (The ATLAS Collaboration), JINST 3 S08003, {\bf 214} (2008). 
\bibitem{alfa} G. Aad {\it et al.}, (The ATLAS Collaboration), JINST 3 S08003, {\bf 211} (2008). 
\bibitem{fp-ref1} B.E. Cox and J.R. Forshaw, Comput. Phys. Commun. 144, 104 (2002).
\bibitem{higgs-cep} V.A. Khoze, A.D. Martin and M.G. Ryskin, Eur. Phys. J. C. {\bf 34} (2004) 327; J.R. Forshaw, arXiv:0901.3040 (2009).
\bibitem{cdfCEP} T. Aaltonen {\it et al.} [CDF Coll.], Phys. Rev. Lett. 99 (2007) 242002;  T. Aaltonen {\it et al.} [CDF Coll.], Phys. Rev. D 77 (2008) 052004.
\bibitem{exhume} J. Monk and A. Pilkington, Comput. Phys. Communication 175 (2006) 232.
\bibitem{afp} S. Heinemeyer {\it et. al.}, Eur. Phys. J. C 53 (2008). 
\bibitem{mssm} B. Cox {\it et. al.}, JHEP 0710:090, (2007). 
\end{thebibliography}
\end{document}